# Enabling Highly Efficient Infrared Silicon Photodetectors via Disordered Metasurfaces with Upconversion Nanoparticles

Wei Chen[1,2,#], Shutao Zhang[2,3,4,#], Chongwu Wang[5,6,#], Yiming Wu[7], Xiaodong Shi[8], Jiaqing Shen[1], Yan Liu[2,8], Xuran Zhang[5,6], Febiana Tjiptoharsono[2], Henry Yit Loong Lee[2,8], Di Zhu[4,8,9], Qijie Wang[5,6], Joel K. W. Yang[3,10, *], Jinfeng Zhu[1, *], Zhaogang Dong[2,3,8, *]

[1]Institute of Electromagnetics and Acoustics and Key Laboratory of Electromagnetic Wave Science and Detection Technology, Xiamen University, Xiamen, Fujian 361005, China

[2]Institute of Materials Research and Engineering (IMRE), Agency for Science, Technology and Research (A*STAR), 2 Fusionopolis Way, Innovis #08-03, Singapore 138634, Republic of Singapore

[3]Singapore University of Technology and Design, 8 Somapah Road, 487372, Singapore

[4]Department of Materials Science and Engineering, National University of Singapore, Singapore 117575, Singapore

[5]School of Electrical and Electronic Engineering, Nanyang Technological University, Singapore 639798, Singapore

[6]CINTRA (CNRS-International-NTU-THALES Research Alliance), IRL 3288, Research Techno Plaza, 50 Nanyang Drive, Border X Block, Level 6, Singapore, 637553 Singapore

[7]Institute of Flexible Electronics (Future Technologies), Xiamen University, Xiamen, 361005, China

[8]Quantum Innovation Centre (Q.InC), Agency for Science Technology and Research (A*STAR), 2 Fusionopolis Way, Innovis #08-03, Singapore 138634, Republic of Singapore

[9]Centre for Quantum Technologies, National University of Singapore, Singapore 117543, Singapore

[10]Singapore-HUJ Alliance for Research and Enterprise (SHARE), The Smart Grippers for Soft Robotics (SGSR) Programme, Campus for Research Excellence and Technological Enterprise (CREATE), Singapore 138602

[#] These authors equally contribute to this work.

*Correspondence and requests for materials should be addressed to

J.K.W.Y. (email: joel_yang@sutd.edu.sg),

J.Z. (email: jfzhu@xmu.edu.cn),

Z.D. (email: Zhaogang_dong@sutd.edu.sg).

ORCID

Zhaogang Dong: http://orcid.org/0000-0002-0929-7723

Joel K. W. Yang: http://orcid.org/0000-0003-3301-1040

Jinfeng Zhu: https://orcid.org/0000-0003-3666-6763

Qijie Wang: https://orcid.org/0000-0002-9910-1455

Di Zhu: https://orcid.org/0000-0003-0210-1860

Wei Chen: https://orcid.org/0000-0003-0103-3149

**ABSTRACT**

Silicon photodetectors are highly desirable for their CMOS compatibility, low cost, and fast response speed. However, their application the infrared (IR) is limited by silicon's intrinsic bandgap, which restricts its detection to photons with wavelengths shorter than 1.1 μm. Although several methods have been developed to extend silicon photodetectors further in the IR range, these approaches often introduce additional challenges, such as increased fabrication complexity and compatibility issues with standard CMOS processes. Here, we present an approach to overcome these limitations by integrating disordered metasurfaces with upconversion nanoparticles (UCNPs), enabling IR detection by silicon photodetectors. The disordered design consists of hybrid Mie-plasmonic cavities, which can enhance both the near-field localization and wide-band light absorption from visible to IR, improving photocurrent conversion. Compared to ordered structures, the infrared absorption and near field of the highly disordered configuration are increased by 2.6-folds and 3.9-folds, respectively. UCNPs not only convert near-infrared photons into visible light but also enhance absorption in the mid-infrared range, thereby improving hot electron generation. The measured responsivity of the disordered element for 1550 nm laser is up to 0.22 A/W at room temperature, corresponding to an external quantum efficiency of 17.6%. Our design not only enhances the photocurrent performance significantly, but also extends the working wavelength of silicon photodetectors to IR wavelength, making them suitable for broad spectrum applications.

**KEYWORDS**: Photodetector, upconversion nanoparticles, disordered metasurface, infrared light

# 1. Introduction

Silicon (Si)-based photodetectors offer notable advantages, including compatibility with CMOS processing, low cost, excellent stability, miniaturization, and high integration potential with other electronic components [1-5]. These attributes make Si chips the preferred platform for the large-scale integration of multiple optical and electronic functions. However, a significant limitation of Si photodetectors stems from their intrinsic properties, particularly its relatively wide bandgap of 1.1 eV [6-7]. This bandgap renders Si optoelectronic devices ineffective at absorbing and detecting infrared (IR) light, thus constraining their applications in the IR wavelength range, such as telecommunications, medical diagnostics, and remote sensing [8-11]. Therefore, extending the operational wavelength range of Si photodetectors beyond their inherent bandgap is crucial for future technological advancements.

Several strategies have been explored to circumvent this limitation, such as the development of Ge or III-V/2D-material hybrid photodetectors, bandgap engineering through doping, and leveraging the internal photoemission effect [12-15]. Unfortunately, these methods increase the complexity of preparation and reduce quantum efficiency [16-17]. On the other hand, integrating Si detectors with upconversion nanoparticles (UCNPs) could provide a promising way for extending the working wavelength to infrared wavelength, as UCNPs can convert low-energy photons into high-energy ones, through the anti-Stokes process [18-20]. However, so far, Si infrared photodetectors with UCNPs remain inefficient due to their inherently low absorption coefficients (responsivity only 8 mA/W) [21]. Thus, enhancing the upconversion efficiency of UCNPs remains a significant challenge for Si-based photodetectors [22-23].

In this work, we present a highly efficient Si-based IR photodetector based on disordered Mie-plasmonic metasurfaces integrated with upconversion nanoparticles (UCNPs), which facilitate the upconversion of multiple infrared photons into visible ones via the anti-Stokes process. Our disordered metasurface design not only confines the electric field to be within the Si nanopillars, but also significantly broadens light absorption across the visible to near-infrared spectrum. As a result, the nanostructured detector demonstrates a cutoff wavelength of 1.55 μm and achieves an external quantum efficiency (EQE) of 17.6%. Finite-difference time-domain (FDTD) simulations reveal that the broadband near-field enhancement originated from the high degree-of-disorder in the metasurface structure. This disordered resonance system introduces a novel approach to photodetection and holds promise for a wide range of optoelectronic applications.

## 2. Results

### 2.1 Silicon IR Detectors based on Disordered Metasurfaces as Integrated with Upconversion Nanoparticles

Figure 1a presents the schematic of our designed Si-based IR detector that utilizes disordered metasurfaces integrated with UCNPs. The UCNPs, comprising $NaYF_4$, $Er@NaYF_4$, are synthesized as detailed in the Methods section, with morphology shown in Fig. S1. These nanoparticles are integrated into the metasurface, enabling the upconversion from incident infrared light into multiple-wavelength visible light. The converted visible photons can be detected via a Schottky barrier between the Al film and the Si substrate in Fig. S2. The enhanced mechanism behind the disordered metasurface integrated with UCNPs is plotted in the right plane. For the IR wavelength, the disordered metasurface enhances light absorption and boosts upconversion efficiency. At visible wavelengths, this structure confines the electric field within the Si nanopillars, facilitating efficient photon absorption and subsequent electron-hole pair generation. Therefore, our disordered metasurfaces integrated with UCNPs significantly improve the photogenerated current, leading to the elevation of detection performance.

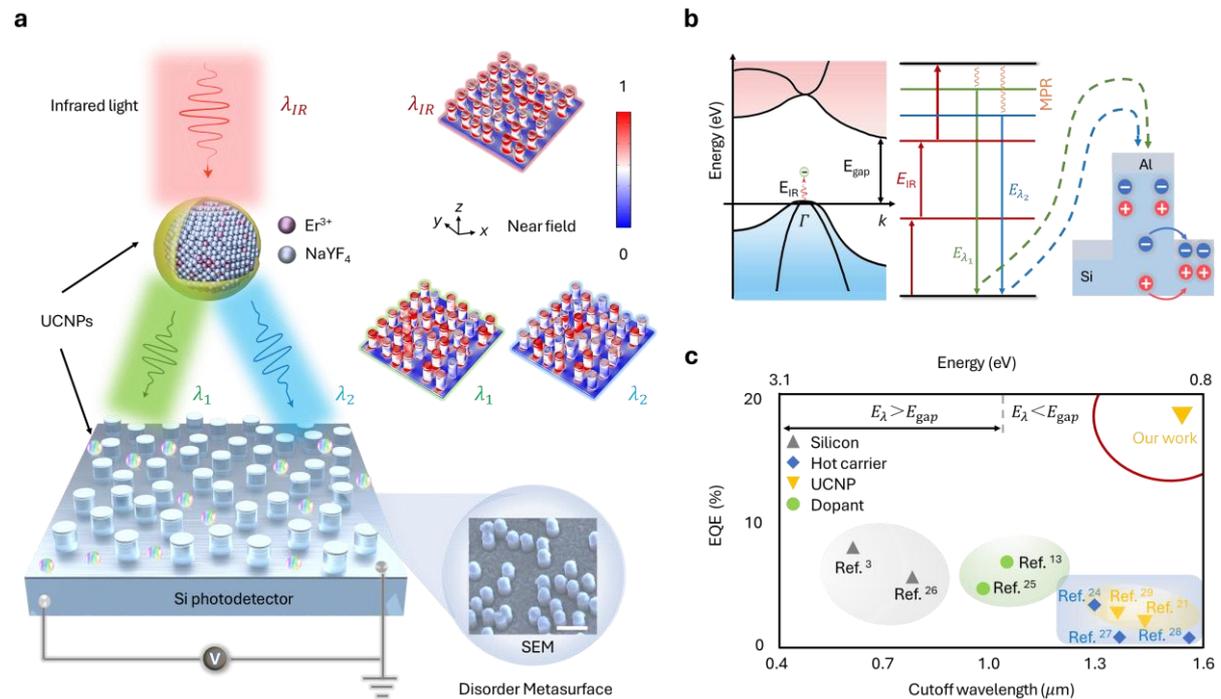

**Figure 1. Schematic of our designed Si near-IR photodetector based on the disordered metasurfaces with UCNPs.** (a) Schematic illustration of the disordered metasurfaces with UCNPs. Disordered metasurface for enhancing light absorption at both visible and infrared wavelengths of $\lambda_1$=550 nm, $\lambda_2$=650 nm, $\lambda_{IR}$=1550 nm and the scale bar of the scanning electron microscopy is 500 nm. (b) Energy level of Si and upconversion principle from infrared to visible light, where MPR denotes multiphoton relaxation. (c) Positioning this work with respect to prior reports of EQE and cutoff wavelengths for Si-based detectors (details in Table S1).

As illustrated in Fig. 1b, the band structure of Si has an energy gap $E_{gap}$ of approximately 1.1 eV, which is larger than the energy of near-IR photons ($E=h\nu$). Therefore, this characteristic limit IR spectrum application for Si-based photodetectors. Based on the anti-Stokes effect of UCNPs, two or multiple infrared photons can be converted into visible photons via the upconversion process. Building on this principle, a benchmark comparison of Si-based photodetectors in Fig. 1c highlights the advantages of our device configuration, underscoring the potential of this design to empower Si-based IR photodetection [24-30]. Compared to existing UCNPs technologies [21], our approach boosts the responsivity by 28-fold.

**2.2 Working Principle of the Disordered Metasurface Cavities**

The inherent imperfections in nanophotonic fabrication and the unique properties of the disorder have motivated extensive exploration into disordered metasurfaces, which has given rise to intriguing phenomena owing to the complex nature of their interaction with light, as compared to perfect periodic metasurfaces [31]. Here, to study the impact of disordered metasurface on Si-based photodetectors, we introduce disorder into the system by varying the positions of the nanopillars, as described by a parameter $\sigma$, which quantifies the degree-of-disorder. The positions of the nanopillars $(x^i, y^i)$ are modified as follows [32],

$$x^i = x_0^i + (\sigma * P_0)U_1(X, Y), \qquad (1)$$

$$y^i = y_0^i + (\sigma * P_0)U_2(X, Y), \qquad (2)$$

where $(x_0^i, y_0^i)$ denotes the position of the nanopillars in the periodic array with periodicity $P_0$, $U_1$ and $U_2$ are two independent uniform distributions within the range $[-1, 1]$. $\sigma$ is a control parameter representing the degree-of-disorder in the system.

For instance, here we set $\sigma=0$, $\sigma=0.3$, and $\sigma=0.7$ to deduce how the increasing disorder influences the performance of the Mie-plasmonic hybrid cavities (see the unit cell in Fig. S2). Full-wave simulations based on the 3D FDTD method are implemented to elucidate the mechanism of disordered metasurface (see Methods). Introducing disorder breaks the translational symmetry and makes the band structure of a photonic cavity not well-defined [33-34]. The resulting spectra exhibit several pronounced peaks, corresponding to spectral gaps associated with either Mie or Bragg resonances [35-36]. The Mie-plasmonic hybrid cavities more effectively trap photons to reduce the corresponding reflection compared to order structures, as the degree-of-disorder increases [37]. Specifically, the degree-of-disorder with $\sigma$=0.7 demonstrates a 162% enhancement in average absorption in the visible band and an 89% improvement in the IR band compared to the bare periodic Si structures with $\sigma$=0 (see $\sigma$=1 in Fig. S3). This dual-band absorption capability enables efficient upconversion and facilitates the generation of photogenerated currents.

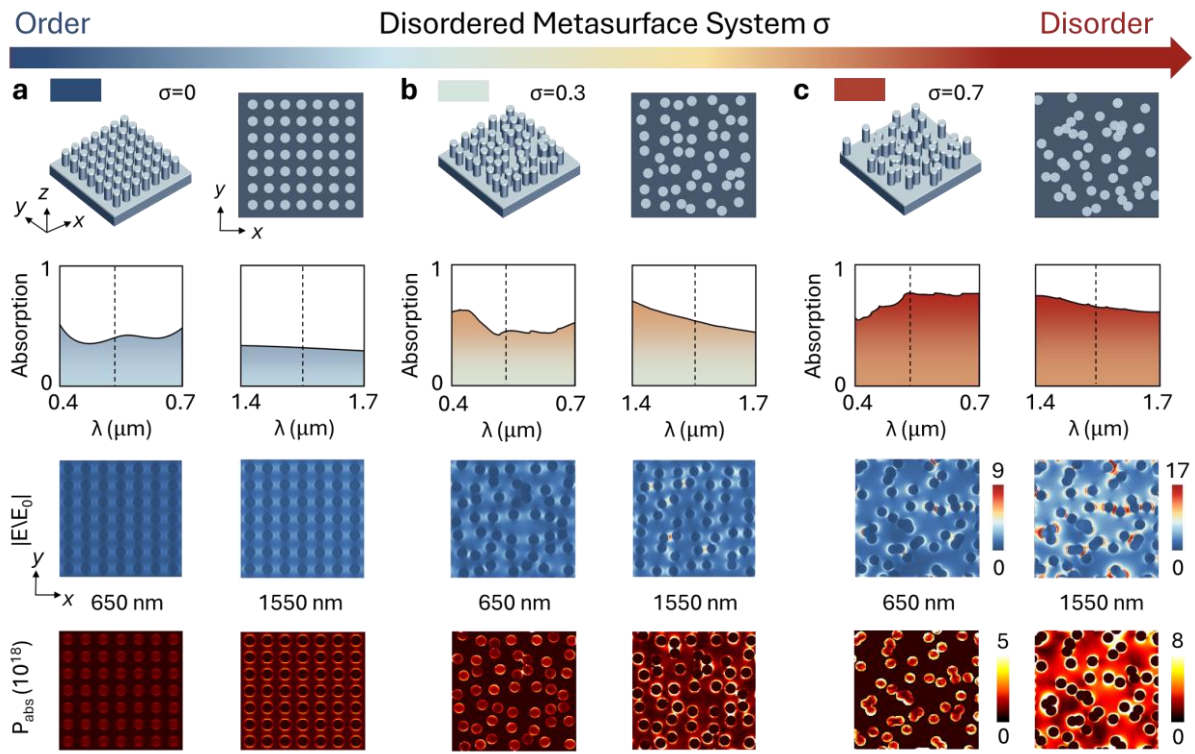

**Figure 2. Disordered design to boost the optical performance of metasurface detectors.** Arrangement of unit cells, absorption spectra, field enhancement $|E/E_0|$, and absorption intensity distribution $P_{abs}$ for (a) $\sigma$=0, (b) $\sigma$=0.3, and (c) $\sigma$=0.7. Wavelengths of 650 nm and 1550 nm are corresponding to the central wavelengths for UCNPs operation in the visible and infrared ranges.

To gain physical insights, we plot the electric field enhancement distributions $|E/E_0|$ for $\lambda$=650 nm and $\lambda$=1550 nm [38]. In the leftmost panel (Fig. 2a), the unit cells are separated by a uniform distance which we call the middle gap, leading to a fixed response with limited broadband capability, as only a single lattice resonance is present at this stage (see Fig. S4) [39]. In Fig. 2b, more combinations are randomly introduced, *e.g.*, long gap, and short gap, leading to various resonances of individual particles as well as lattice resonances at different periods. The rightmost panel (Fig. 2c) reveals that a high degree-of-disorder further introduces overlap, enhancing the coupling effect between neighboring particles [40-42]. As a result, the intensity and distribution of the near field are significantly altered by the increasing degree of disorder. In addition, absorption intensity distribution $P_{abs}$ highlights the distinct contributions of disordered metasurfaces in both visible and IR wavelengths [43-44]. In the visible spectrum, the Si nanopillars primarily dominate absorption, while in the IR band, most of the photon absorption occurs in the aluminum nanostructures (see 3D $P_{abs}$ in Fig. S5). Compared to the ordered structure, the infrared $P_{abs}$ and |E/E$_0$| of the highly disordered configuration increased by 2.6 and 3.9 folds, respectively. A series of systematic investigations imply that the disordered design has the potential to boost the optical performance of metasurfaces significantly, including light absorption enhancement, near-field localization, and absorbed power elevation [45-46].

**2.3 Experiment of Disordered Metasurfaces with Upconversion Nanoparticles**

To test this principle, we fabricate two disordered detector samples using electron-beam lithography, one with and another without UCNPs (see Methods). As shown in Fig. 3a, the optical microscope image of the disordered metasurfaces appears black with a slight blue hue. The scanning electron microscopy (SEM) image highlights the disordered feature, with nanopillars positioned at various distances, including far, close, and overlapping arrangements (see more details in Fig. S6). After spin-coating UCNPs (NaYF$_4$, Er@NaYF$_4$) onto the prepared disordered metasurface, a visible color change from dark blue to dark red is observed.

The SEM in Fig. 3b shows the presence of UCNPs with a size of approximately 20 nm around the nanopillars [47-48]. The absorption spectra before and after depositing UCNPs are shown in Fig. 3c. As can be seen, the metasurface structure we designed exhibits a significant enhancement in absorption compared to the flat region, both in the visible and infrared (IR) wavelengths. More importantly, after the attachment of UCNPs, the metasurface demonstrates superior and selective absorption in the IR wavelengths, which is highly beneficial for photodetection experiments that rely on photon absorption. In the mid-infrared region, the absorption peaks at 3.5–3.6 µm correspond to the symmetric and asymmetric stretching vibrations of the methylene group in oleic acid molecules, confirming successful attachment (see Method 4.3 Synthesis of upconversion nanoparticles) [49-50].

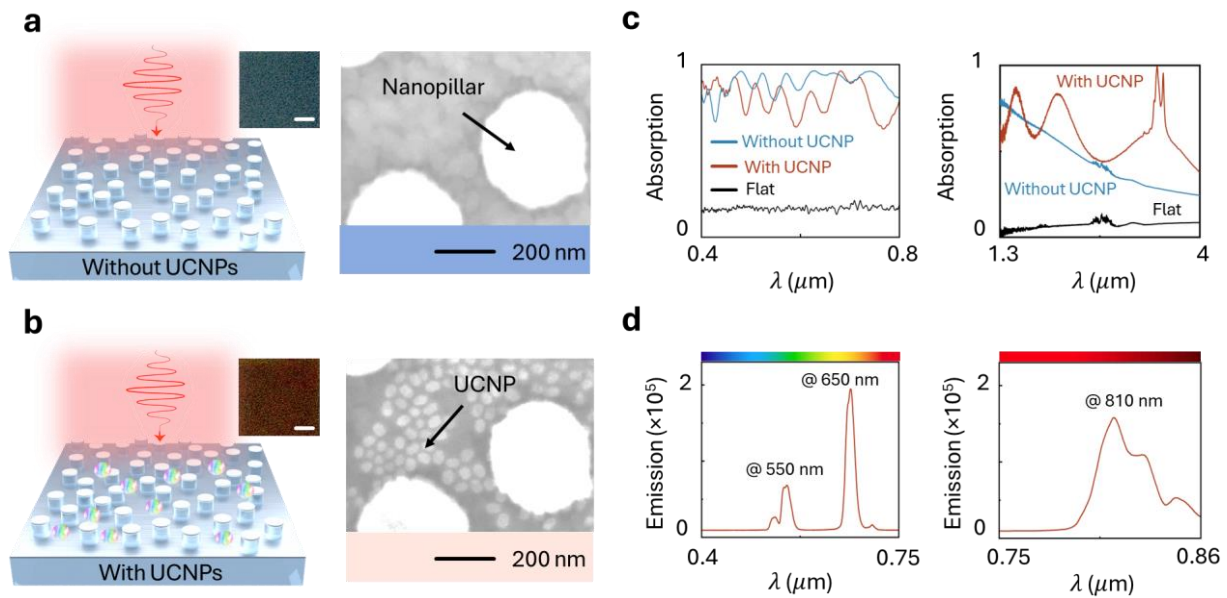

**Figure 3. Converting infrared light into visible light by disordered metasurfaces with UCNPs.** Schematic and SEM for the disordered metasurface (a) without and (b) with UCNPs. Insets are the bright-field optical microscope images with a scale bar of 50 μm. (c) Measured absorption spectra at visible and IR wavelengths. (d) Plot of upconversion emission (a. u.) at different wavelength for the disordered metasurface with a 1550-nm laser.

Using a custom-built optical microscope shown in Fig. S7, we can characterize the intrinsic luminescence spectra from the disordered metasurfaces with UCNPs, where Fig. 3d demonstrates a significant intensity enhancement at two main emission bands: ~550 nm and ~650 nm. In contrast, for the disordered metasurface without UCNPs, the luminescence intensity is zero across the entire spectrum, indicating no upconversion behavior from infrared to visible photon (see Fig. S8). As expected, due to the bandgap limitations, the Si-based metasurface has no response to IR irradiation. After dispersing the UCNPs, the metasurface converts infrared photons of 1550 nm into higher-energy visible photons within the range of 500-820 nm ($E = h\nu$), as shown in Fig. 3d. More importantly, this metasurface exhibits wavelength selectivity and strong absorption in these wavelengths, which is expected to greatly enhance the photocurrent responsivity.

## 2.4 Realization of Si-based IR Photodetector via Our Design

To demonstrate the feasibility of our disordered metasurface with UCNPs for Si-based IR photodetectors, we set up a measurement system as plotted in Fig. 4a. The disordered metasurface samples are fixed on the printed circuit board (PCB), for stable wiring, enabling photocurrent measurements by providing electrical connections. where the left side is for voltage biasing, and the right-side ports are for current measurements.. For benchmarking investigations, we use two identical disordered metasurface samples with and without UCNPs. Fig. 4b illustrates that, under the illumination of a 1550 nm continuous-wave (CW) laser with an incident power of 0.4 mW, the photogenerated current of the disordered metasurface without UCNPs is always zero regardless of the bias voltage. In comparison, the sample with UCNPs is able to exhibit a remarkable photogenerated current of up to 88 µA at a reverse bias voltage of 20 V. This performance demonstrates that our design effectively extends the functionality of Si-based photodetector into the IR band, overcoming the inherent energy band limitation of Si (see dark current in Fig. S9).

We also compare the *I-V* characteristics as measured from the flat and disordered metasurface regions as shown in Fig. 4c, where the disordered metasurface enhances the photogenerated current by 19-fold at a reverse bias voltage of 20 V. To further verify the robustness, we measure the current at three random points within the disordered metasurface region. The results in Fig. 4d show great agreement under a voltage switch from 0 to 20 V, illustrating the generality and stability of our design. We next investigate the dependence of *I-V* responses of two disordered metasurface samples under different laser powers in Figs. 4e. While the device without UCNPs remains inactive regardless of power changes, the sample with UCNPs exhibits a clear dependence of *I-V* response on the power (see Fig. S10). Under 1550 nm NIR light illumination at a pump power of 0.4 mW, the measured photocurrent is 0.088 mA, corresponding to a responsivity of 0.22 A/W and EQE of 17.6% (see Supporting Information). This impressive photovoltaic performance arises from the upconversion process and broadband resonance enhancement enabled by the disordered metasurfaces [51-52].

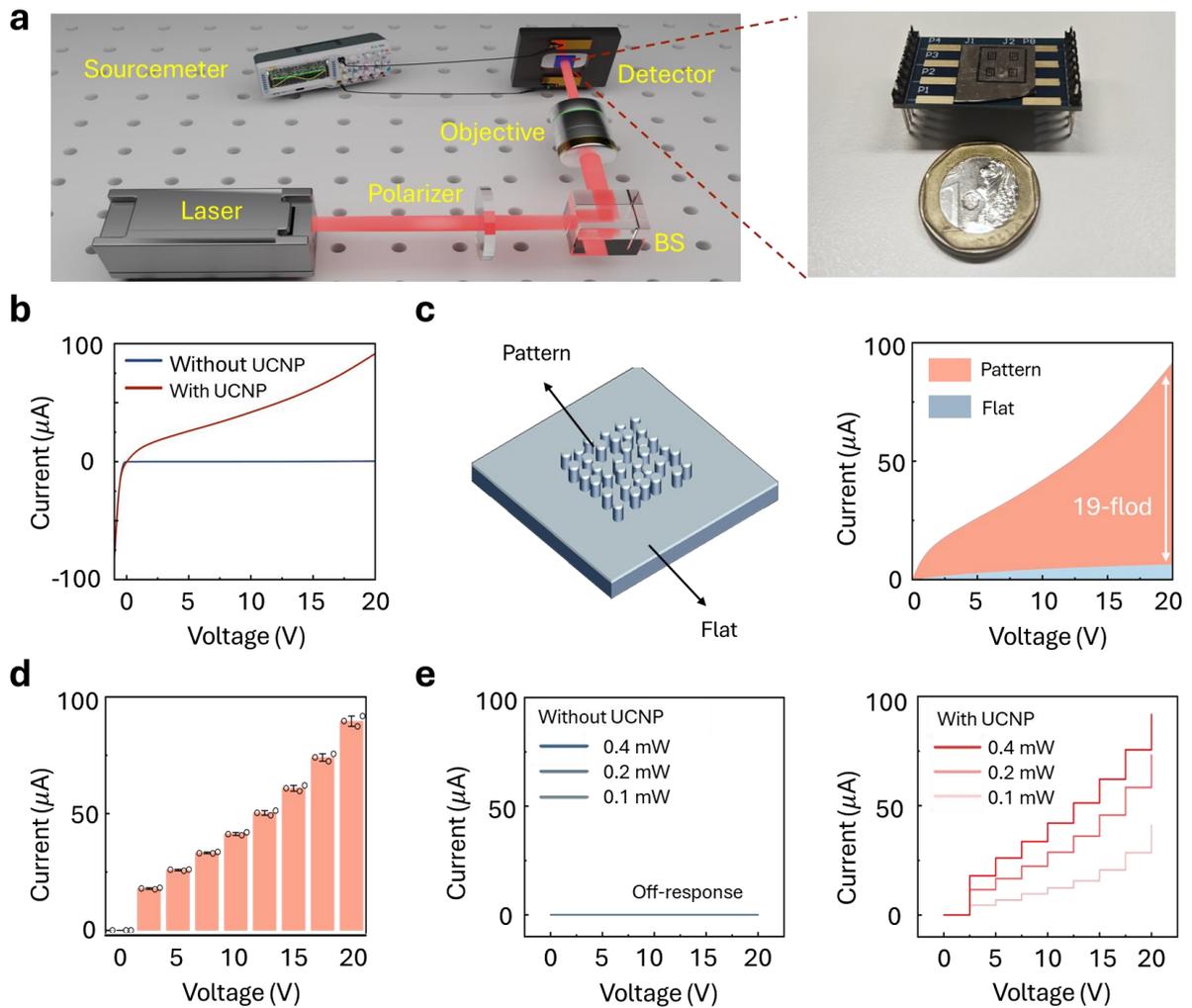

**Figure 4. Room-temperature measurement of our proposed Si-based IR photodetector illuminated with 1.55 μm-laser.** (a) Schematic of the experimental setup for optoelectronic response and a photograph of our detector with an SGD coin. (b) Comparison of *I-V* curve testing for disordered metasurfaces with and without UCNPs. (c) Disordered-metasurface-enhanced photocurrent intensity compared to that of the flat region. (d) *I-V* characteristics for randomly selected points in the pattern region of the device. (e) Comparisons on the photocurrent responses under different CW laser powers for two disordered metasurface samples.

## 3. Conclusions

In this paper, we design a silicon-based photodetector based on disordered metasurfaces as integrated with UCNPs, where its operation wavelength can be extended into the near-IR and mid-IR range. Our disordered metasurface, analyzed by full-wave optical simulation, exhibits enhanced near-field localization capabilities, significantly improving the absorption in both visible and IR bands. For the near-IR region, the incorporation of UCNPs enables the upconversion of infrared photons into visible photons, which are effectively detected by silicon photodetectors. Our room-temperature experimental results reveal that the disordered metasurface-UCNPs hybrid system can generate a considerable photocurrent in the infrared region, with the disordered metasurface design dramatically boosting the quantum efficiency. These findings provide a promising path toward next-generation, broadband silicon-based optoelectronic devices, such as the miniaturized IR spectrometers with the potential integration with tunable metasurface configurations [53-55].

## 4. Methods

### 4.1 Numerical simulations.

All numerical simulations are performed using the Finite-Difference Time Domain (FDTD) Solutions, ANSYS Inc. software. The refractive index of Aluminum (Al) and silicon (Si) is sourced from the Handbook of Chemistry and Physics and plane waves propagate along the z-axis [56]. For the periodic array of Al nanopillars, we simulate a 7×7 array considering the limitations of the computing devices [31]. The period between each nanopillar is 310 nm.

### 4.2 Fabrication of the disordered metasurface.

The hydrogen silsesquioxane (HSQ) etching mask was prepared on a single-crystalline *p*-type doped Si wafer (resistivity: 10-50 ohm·cm, Slicon Valley Microelectronics Inc.). HSQ resist (Dow Corning, XR-1541-006) was diluted to 3% using methyl isobutyl ketone (MIBK) and then spin-coated on the cleaned substrate at 3000 rpm, resulting in a film thickness of approximately 50 nm. Electron beam lithography (Elionix ELS-7000) was performed at an acceleration voltage of 100 keV, with a beam current of 500 pA and a dose of ~12 mC/cm². The sample was developed in a NaOH/NaCl solution (1 wt %/4 wt % in deionized water) for 60 seconds, followed by immersion in deionized water for 60 seconds to stop the process. Afterward, the sample was rinsed with acetone and isopropanol and dried under a continuous nitrogen flow. Silicon etching to a depth of 200 nm was performed using inductively coupled plasma etching (Oxford Instruments, Plasmalab System 100) under 100 W radio frequency power, 150 W ICP power, with $Cl_2$ gas at a flow rate of 22 sccm, at a pressure of 5 mTorr and a temperature of 40°C. A 40 nm-thick aluminum layer was then deposited onto the silicon nanostructures via electrical beam evaporation.

**4.3 Synthesis of upconversion nanoparticles.**

$NaYF_4$: Er (10%) core nanoparticles were synthesized using a co-precipitation method. Typically, a 2-mL aqueous solution of Ln $(CH_3CO_2)_3$ (0.2 M, Ln= Y, and Er) was added to a 50-mL flask containing 3 mL of oleic acid and 7 ml of 1-octadecene. The reaction mixture was heated to 150 °C and kept for one hour under stirring to remove water from the solution. After cooling to room temperature, a 6-mL methanol solution containing $NH_4F$ (1.6 mmol) and NaOH (1 mmol) was added to the mixture and kept under stirring for 30 min. After removing methanol, the solution was heated at 290 °C under argon for 2 hours and then cooled to room temperature. The resulting nanoparticles were washed with ethanol several times and re-dispersed in 2 mL of cyclohexane.

### 4.4 Synthesis of NaYF4: Er (10%)@NaYF4 core-shell nanoparticles.

The coating procedure of NaYF$_4$ shell is similar to that of core nanoparticles described above, except those aqueous solutions of 2-mL of Y(CH$_3$CO$_2$)$_3$ (0.2 M) together with core nanoparticles dispersed in cyclohexane, and a methanol solution of NH$_4$F and NaOH were used to initiate an epitaxial shell growth process. The resulting core@shell nanoparticles were collected by centrifugation, washed with ethanol, and re-dispersed in 2 mL of cyclohexane.

### 4.5 Morphology characterization.

Reflectance measurements were conducted using a polarized broadband light source and a CRAIC UV-VIS-NIR micro-spectrophotometer, equipped with a ×5 objective lens and a numerical aperture (NA) of 0.12. The system was calibrated using a NISR standard sample (CRAIC Technologies) to ensure absolute reflectance values. The measurements were recorded with a 100 ms integration time, averaging 50 spectra per sample. SEM images are taken at an acceleration voltage of 1 keV (Hitachi, SU8220).

### 4.6 Photoluminescence measurements.

Photoluminescence measurements are carried out using a customized Olympus BX51 microscope, integrated with a 1550-nm continuous-wave laser. The excitation laser at 1550 nm was defocused through a 100× objective lens (0.9 NA) to create a 15-μm diameter spot on the sample. The emitted luminescence signal was captured using an Ocean Optics QR Pro spectrometer.

### 4.7 Photocurrent measurements with near-IR and mid-IR lasers.

A continuous-wave laser (Santec TLS-570) is used as the pump to stimulate the Si-based detector, and its wavelength is set to 1550 nm [57]. Our metasurface unit consists of Si nanopillars, Al pillars-on top, and an Al film, with a *p*-type Si substrate. The holes in the *p*-Si region diffuse towards the Al to equilibrate the Fermi level, forming a Schottky barrier between the Al film and the *p*-type Si substrate. As a result, the Al film becomes positively charged while the *p*-Si region acquires a negative charge, creating a depletion region. The electrically modulated photoluminescence (PL) spectra were obtained using a computer-controlled dual-channel sourcemeter (Keithley 2636B) to simultaneously apply DC voltage and measure the device's electric current, combined with a WITec Alpha 300 S Scanning Near-field optical microscope.

**Notes**

The authors declare no competing financial interests.

**Author contributions**

Z.D. and J.K.W.Y. conceived the concepts and supervised the project. Z.D. and J.Z. conceived the concept of a Si-based infrared detector. S.Z., W.C., F.T. and Z.D. prepared the metasurface samples. Y.W. prepared the upconversion nanoparticle samples. W.C., and H.L.Y.L. did the SEM characterizations. Y.L. did the back focal plane measurements at the visible wavelength. J.S., C.W., X.Z., and Q.W. did the mid-IR device characterizations and interpretations. W.C. and Y.L. did the numerical simulations. S.Z. and W.C. did the optical reflectance measurements. C.W., X. Z., X.S. and D.Z. set up the near-IR measurement system at 1.55 µm. J.Z. participated in the discussions and provided suggestions. The paper was drafted by W.C., with input from Z.D., J.K.W.Y. and J.Z. All authors analyzed the data and read and corrected the manuscript before the submission. W.C. and S.Z. are equal contributions.

**Acknowledgments**

Z.D. and J.K.W.Y. would like to acknowledge the funding support from The National Research Foundation (NRF), Singapore via Grant No. NRF-CRP30-2023-0003. In addition, Z.D. would like to acknowledge the funding support from the Agency for Science, Technology and Research (A*STAR) under MTC IRG (Project No. M21K2c0116 and M22K2c0088), and the Quantum Engineering Program 2.0 (Award No. NRF2021-QEP2-03-P09). J.Z. would like to acknowledge the funding support from NSFC (62175205), NSAF (U2130112), Natural Science Foundation of Fujian Province (2024J02005), the Youth Talent Support Program of Fujian Province (Eyas Plan of Fujian Province) [2022] and Shenzhen Science and Technology Development Funds (Grant No. JCYJ20220530143015035). J.K.W.Y. would like to acknowledge the funding support from National Research Funding (NRF) Singapore NRF-CRP20-2017-0001 and NRF-NRFI06-2020-0005. In addition, W.C. would like to acknowledge the funding supporting from the China Scholarship Council Scholarship (CSC NO. 202306310153). J.S. would like to acknowledge the funding supporting from the China Scholarship Council Scholarship (CSC NO. 202306310168). This research is also supported by grants from the National Research Foundation, Prime Minister's Office, Singapore under its Campus of Research Excellence and Technological Enterprise (CREATE) programme. Y.W. would like to acknowledge the funding support from National Natural Science Foundation of China (Grant No. 52372156 and 62288102).

**References**

[1] Liu, Chaoyue, et al. "Silicon/2D-material photodetectors: from near-infrared to mid-infrared." Light: Science & Applications 10.1 (2021): 123.

[2] Michel, Jurgen, Jifeng Liu, and Lionel C. Kimerling. "High-performance Ge-on-Si photodetectors." Nature photonics 4, no. 8 (2010): 527-534.

[3] Ho, Jinfa, et al. "Miniaturizing color-sensitive photodetectors via hybrid nanoantennas toward submicrometer dimensions." Science Advances 8.47 (2022): eadd3868.

[4] Kodigala, Ashok, et al. "High-performance silicon photonic single-sideband modulators for cold-atom interferometry." Science Advances 10.28 (2024): eade4454.

[5] Yu, Ting, et al. "Graphene coupled with silicon quantum dots for high-performance bulk-silicon-based Schottky-junction photodetectors." Advanced Materials 28.24 (2016): 4912-4919.

[6] Wang, Mao, et al. "Silicon-based intermediate-band infrared photodetector realized by Te hyperdoping." Advanced Optical Materials 9.4 (2021): 2001546.

[7] Casalino, M., et al. "Near-Infrared All-Silicon Photodetectors." International Journal of Photoenergy 2012.1 (2012): 139278.

[8] Jiang, Hao, et al. "Synergistic-potential engineering enables high-efficiency graphene photodetectors for near-to mid-infrared light." Nature Communications 15.1 (2024): 1225.

[9] Valencia Molina, Laura, et al. "Enhanced Infrared Vision by Nonlinear Up-Conversion in Nonlocal Metasurfaces." Advanced Materials 36.31 (2024): 2402777.

[10] Dai, Mingjin, et al. "Long-wave infrared photothermoelectric detectors with ultrahigh polarization sensitivity." Nature Communications 14.1 (2023): 3421.

[11] Dai, Mingjin, et al. "On-chip mid-infrared photothermoelectric detectors for full-Stokes detection." Nature Communications 13.1 (2022): 4560.

[12] Yin, Yanlong, et al. "High-speed and high-responsivity hybrid silicon/black-phosphorus waveguide photodetectors at 2 µm." Laser & Photonics Reviews 13.6 (2019): 1900032.

[13] Li, Tengfei, et al. "Sensitive photodetection below silicon bandgap using quinoid-capped organic semiconductors." Science Advances 9.13 (2023): eadf6152.

[14] Fu, Jintao, et al. "Schottky infrared detectors with optically tunable barriers beyond the internal photoemission limit." The Innovation 5.3 (2024).

[15] Gao, Yifan, et al. "High responsivity p-GaSe/n-Si van der Waals heterojunction phototransistor with a Schottky barrier collector for ultraviolet to near-infrared band detection." Applied Physics Letters 123.8 (2023).

[16] Zhang, Shutao, et al. "Chalcogenide Metasurfaces Enabling Ultra-Wideband Detectors from Visible to Mid-infrared." arXiv preprint arXiv:2409.04763 (2024).

[17] Fu, jintao, et al. "Photo-driven fin field-effect transistors." Opto-Electronic Science 3.5 (2024): 230046-1.

[18] Schiattarella, Chiara, et al. "Directive giant upconversion by supercritical bound states in the continuum." Nature 626.8000 (2024): 765-771.

[19] Huang, Ling, and Gang Han. "Triplet-triplet annihilation photon upconversion-mediated photochemical reactions." Nature Reviews Chemistry 8.4 (2024): 238-255.

[20] Chen, Mian, et al. "Upconversion dual-photosensitizer-expressing bacteria for near-infrared monochromatically excitable synergistic phototherapy." Science Advances 10.10 (2024): eadk9485.

[21] Xiang, Hengyang, et al. "Upconversion nanoparticles extending the spectral sensitivity of silicon photodetectors to λ= 1.5 μm." Nanotechnology 31.49 (2020): 495201.

[22] Han, Sanyang, et al. "Photon upconversion through triplet exciton-mediated energy relay." Nature Communications 12.1 (2021): 3704.

[23] Wu, Yiming, et al. "Upconversion superburst with sub-2 μs lifetime." Nature nanotechnology 14.12 (2019): 1110-1115.

[24] Frydendahl, Christian, et al. "Giant enhancement of silicon plasmonic shortwave infrared photodetection using nanoscale self-organized metallic films." Optica 7.5 (2020): 371-379.

[25] Geng, Xiangshun, et al. "Ultrafast photodetector by integrating perovskite directly on silicon wafer." ACS nano 14.3 (2020): 2860-2868.

[26] Desiatov, Boris, and Marko Lončar. "Silicon photodetector for integrated lithium niobate photonics." Applied Physics Letters 115.12 (2019).

[27] Sobhani, Ali, et al. "Narrowband photodetection in the near-infrared with a plasmon-induced hot electron device." Nature communications 4.1 (2013): 1643.

[28] Dong, Yajin, et al. "CMOS-compatible broad-band hot carrier photodetection with Cu–silicon nanojunctions." ACS Photonics 9.11 (2022): 3705-3711.

[29] Xiang, Hengyang, et al. "Heavy-metal-free flexible hybrid polymer-nanocrystal photodetectors sensitive to 1.5 μm wavelength." ACS applied materials & interfaces 11.45 (2019): 42571-42579.

[30] Yoo, Tae Jin, et al. "A facile method for improving detectivity of graphene/p-type silicon heterojunction photodetector." Laser & Photonics Reviews 15.8 (2021): 2000557.

[31] Xu, Mingfeng, et al. "Emerging long-range order from a freeform disordered metasurface." Advanced Materials 34.12 (2022): 2108709.

[32] Mao, Peng, et al. "Manipulating disordered plasmonic systems by external cavity with transition from broadband absorption to reconfigurable reflection." Nature communications 11.1 (2020): 1538.

[33] Maguid, Elhanan, et al. "Disorder-induced optical transition from spin Hall to random Rashba effect." Science 358.6369 (2017): 1411-1415.

[34] Hu, Zixian, Changxu Liu, and Guixin Li. "Disordered optical metasurfaces: from light manipulation to energy harvesting." Advances in Physics: X 8.1 (2023): 2234136.

[35] Liu, Changxu, et al. "Disorder-immune photonics based on Mie-resonant dielectric metamaterials." Physical review letters 123.16 (2019): 163901.

[36] Jang, Mooseok, et al. "Wavefront shaping with disorder-engineered metasurfaces." Nature photonics 12.2 (2018): 84-90.

[37] Liu, Changxu, et al. "Disorder-induced topological state transition in photonic metamaterials." Physical review letters 119.18 (2017): 183901.

[38] Gao, Yuan, et al. "Meta-Attention Deep Learning for Smart Development of Metasurface Sensors." Advanced Science (2024): 2405750.

[39] Mao, Peng, et al. "Disorder-induced material-insensitive optical response in plasmonic nanostructures: vibrant structural colors from noble metals." Advanced Materials 33.23 (2021): 2007623.

[40] Yu, Zhipeng, et al. "High-security learning-based optical encryption assisted by disordered metasurface." Nature communications 15.1 (2024): 2607.

[41] Zhu, Sheng-ke, et al. "Harnessing disordered photonics via multi-task learning towards intelligent four-dimensional light field sensors." PhotoniX 4.1 (2023): 26.

[42] Yu, Sunkyu, et al. "Engineered disorder in photonics." Nature Reviews Materials 6.3 (2021): 226-243.

[43] Dong, Zhaogang, et al. "Printing beyond sRGB color gamut by mimicking silicon nanostructures in free-space." Nano letters 17.12 (2017): 7620-7628.

[44] Chen, Wei, et al. "All-Dielectric SERS Metasurface with Strong Coupling Quasi-BIC Energized by Transformer-Based Deep Learning." Advanced Optical Materials 12.4 (2024): 2301697.

[45] Dong, Zhaogang, et al. "Silicon nanoantenna mix arrays for a trifecta of quantum emitter enhancements." Nano Letters 21.11 (2021): 4853-4860.

[46] Chen, Wei, et al. "Broadband solar metamaterial absorbers empowered by transformer-based deep learning." Advanced Science 10.13 (2023): 2206718.

[47] Xu, Jiahui, et al. "Multi-level upconversion polarization enabled by programmable plasmons." Chem 10.2 (2024): 544-556.22

[48] Wu, Yiming, et al. "Dynamic upconversion multicolour editing enabled by molecule-assisted opto-electrochemical modulation." Nature Communications 12.1 (2021): 2022.

[49] Tawfik, Salah M., et al. "Naturally modified nonionic alginate functionalized upconversion nanoparticles for the highly efficient targeted pH-responsive drug delivery and enhancement of NIR-imaging." Journal of Industrial and Engineering Chemistry 57 (2018): 424-435.

[50] Suresh, K., et al. "Energy transfer between optically trapped single ligand-free upconversion nanoparticle and dye." Nanotechnology 34.17 (2023): 175702.

[51] Ji, Yanan, et al. "Huge upconversion luminescence enhancement by a cascade optical field modulation strategy facilitating selective multispectral narrow-band near-infrared photodetection." Light: Science & Applications 9.1 (2020): 184.

[52] Xu, Jiahui, et al. "Multiphoton upconversion enhanced by deep subwavelength near-field confinement." Nano Letters 21.7 (2021): 3044-3051.

[51] Liang, Liangliang, et al. "Incoherent broadband mid-infrared detection with lanthanide nanotransducers." Nature Photonics 16.10 (2022): 712-717.

[52] Lu, Li, et al. "Reversible tuning of Mie resonances in the visible spectrum." ACS nano 15.12 (2021): 19722-19732.

[53] Yang, Zongyin, et al. "Miniaturization of optical spectrometers." Science 371.6528 (2021): eabe0722.

[54] Yoon, Hoon Hahn, et al. "Miniaturized spectrometers with a tunable van der Waals junction." Science 378.6617 (2022): 296-299.

[55] Zhang, Shutao, et al. "Reversible electrical switching of nanostructural color pixels." Nanophotonics 12.8 (2023): 1387-1395.

[56] Haynes, William M. CRC handbook of chemistry and physics. CRC press, 2016.

[57] Shi, Xiaodong, et al. "Efficient photon-pair generation in layer-poled lithium niobate nanophotonic waveguides." Light: Science & Applications 13.1 (2024): 282.